\documentclass[final,5p,times]{elsarticle}

\pdfoutput=1

\usepackage{amsmath}
\usepackage{amssymb}
\usepackage{graphicx}

\journal{Physics Letters B}

\def\nslash{n\!\!\!\slash}
\def\bnslash{\bar n\!\!\!\slash}

\newcommand{\nn}{\nonumber} 
\newcommand{\bn}{{\bar n}}

\newcommand{\SCETa}{\mbox{${\rm SCET}_{\rm I}$ }}
\newcommand{\SCETb}{\mbox{${\rm SCET}_{\rm II}$ }}

\newcommand{\vect}[1]{\mathbf{#1}}
\newcommand{\abs}[1]{\left\lvert #1\right\rvert}
\newcommand{\bra}[1]{\left\langle #1\right\rvert}
\newcommand{\ket}[1]{\left\lvert #1\right\rangle}

\newcommand{\Lqcd}{\Lambda_{\text{QCD}}}
\newcommand{\ee}{\mathrm{e}}
\newcommand{\eUV}{\epsilon_{\text{UV}}}
\newcommand{\eIR}{\epsilon_{\text{IR}}}
\newcommand{\doubleint}{\int\!\!\!\!\!\int}

\newcommand{\eq}[1]{Eq.~\eqref{#1}}
\newcommand{\eqs}[2]{Eqs.~\eqref{#1} and \eqref{#2}}
\newcommand{\fig}[1]{Fig.~\ref{#1}}

\DeclareMathOperator{\Tr}{Tr}

\newcommand{\CF}{C_F}

\newcommand{\taun}{\tau_a^n}

\newcommand{\as}{\alpha_s}

\begin{document}



\begin{frontmatter}

\title{\textbf{Infrared Safety in Factorized Hard Scattering Cross-Sections}}

\author{Andrew Hornig}
\ead{ahornig@berkeley.edu}
\author{Christopher Lee}
\ead{clee@berkeley.edu}
\author{Grigory Ovanesyan}
\ead{ovanesyan@berkeley.edu}
\address{Ernest Orlando Lawrence Berkeley National Laboratory and
University of California, Berkeley, CA 94720, USA}

\date{\today}


\begin{abstract}
The rules of soft-collinear effective theory can be used na\"{\i}vely to write hard scattering cross-sections as convolutions of separate hard, jet, and soft functions. One condition required to guarantee the validity of such a factorization is the infrared safety of these functions in perturbation theory. Using $e^+ e^-$ angularity distributions as an example, we propose and illustrate an intuitive method to test this infrared safety at one loop. We look for regions of integration in the sum of Feynman diagrams contributing to the jet and soft functions where the integrals become infrared divergent. Our analysis is independent of an explicit infrared regulator, clarifies how to distinguish infrared and ultraviolet singularities  in pure dimensional regularization, and demonstrates the necessity of taking zero-bins into account to obtain infrared-safe jet  functions.
\end{abstract}


\end{frontmatter}

\section{Introduction}

Factorization restores predictive power to calculations in Quantum Chromodynamics (QCD) which cannot be carried out exactly due to the contributions of nonperturbative effects. By separating perturbatively-calculable and nonperturbative contributions to observables in QCD and relating the nonperturbative contributions to different observables to each other, we gain the ability to make real predictions. 

Proving factorization rigorously is a technically challenging undertaking, which  traditionally has been formulated in full QCD \cite{Collins:1989gx,Sterman:1995fz}. More recently, many formal elements of these factorization proofs, such as power counting, gauge invariance, the organization of soft gluons into eikonal Wilson lines, and their decoupling from collinear modes, have been organized in the framework of soft-collinear effective theory (SCET) \cite{Bauer:2000ew,Bauer:2000yr,Bauer:2001ct,Bauer:2001yt}. These generic properties of the effective theory allow one to write at least nominally a formula ``factorized'' into collinear (jet) and soft functions for an arbitrary hard scattering cross-section in which strongly-interacting light-like particles participate \cite{Bauer:2002nz}. Examples are the factorization of a large class of two-jet event shape distributions in $e^+e^-$ annihilations to light quark jets \cite{Bauer:2002ie,Bauer:2003di,Bauer:2008dt}, jet mass distributions for $e^+ e^-$ to top quark jets \cite{Fleming:2007qr},  or arbitrary jet cross-sections in $pp$ collisions independently of the choice of actual jet algorithm or observable \cite{Bauer:2008jx}. While the formalism of SCET leads directly to expressing these observables as  convolutions of separate hard, jet, and soft functions, blind use of this procedure without considering further specific properties of each chosen observable can hide whether factorization truly holds in a particular case or not.

An ideal set of observables for which to examine factorizability is the set of angularities $\tau_a$ \cite{Berger:2003iw}, which are two-jet $e^+e^-$ event shapes dependent on a tunable parameter $a$ controlling how sensitive the event shape is to radiation along the jet axes or at wider angles.  Varying $a$ between 0 and 1 interpolates between the thrust \cite{Brandt:1964sa,Farhi:1977sg} and jet broadening \cite{Catani:1992jc} event shapes, but $a$ can take any value $-\infty<a<2$ and give an infrared-safe observable in QCD. Angularities are known to be factorizable, however, only for $a<1$  \cite{Berger:2003iw}. For events $e^+ e^-\rightarrow X$, the angularity of a final state $X$ is
\begin{equation}
\label{tauadef}
\begin{split}
\tau_a(X) = \frac{1}{Q}\sum_{i\in X} E_i\sin^a\theta_i (1-\cos\theta_i)^{1-a} = \frac{1}{Q}\sum_{i\in X} \abs{\vect{p}_i^\perp} \ee^{-\abs{\eta_i}(1-a)}\,,
\end{split}
\end{equation}
where in the first form $E_i$ is the energy of particle $i$ and $\theta_i$ is the angle between its momentum and the thrust axis of $X$. In the second form, $\vect{p}_i^\perp$ is the momentum of particle $i$ transverse to the thrust axis, and $\eta_i $ is its rapidity with respect to the direction of the thrust axis. We assume all final-state particles are massless.

In a separate publication, using SCET, we calculate  the angularity jet and soft functions to next-to-leading order in the strong coupling $\as$,  resum large logarithms using renormalization group evolution, and model the nonperturbative soft function in a way that avoids renormalon ambiguities  \cite{long}.

In this Letter, using angularity distributions as an example, we describe a simple, intuitive method for testing the validity of a factorization theorem deduced from the simple rules of SCET. We begin by na\"{\i}vely presuming the factorizability of a given observable and then attempt to calculate perturbatively the one-loop jet and soft functions. If the factorization holds, each of these functions should be infrared-safe. If they are not, we learn immediately that the factorization breaks down.

Perturbative infrared-safety of jet and soft functions is not, of course, by itself sufficient to guarantee validity of the proposed factorization theorem. The size of power corrections must also be taken into account. The methods we describe in this Letter address only the former issue, not the latter. (Power corrections for angularity distributions and their implications for factorizability were studied in  \cite{Bauer:2008dt,Berger:2003iw,Lee:2006nr}.) However, our method is a quick and direct way to narrow down the class of observables for which a ``generic'' factorization deduced from SCET  (e.g. \cite{Bauer:2008jx}) could actually be valid. 

Our analysis also sheds light on some issues related to infrared divergences in effective theory loop integrals.  Finding a tractable regulator in SCET that suitably controls all infrared divergences has been very challenging (see, e.g., \cite{Bauer:2003td,Chiu:2009yx}). Care is also required to define the effective theory such that it avoids double-counting momentum regions and infrared divergences of full theory diagrams. The procedure of zero-bin subtraction has been proposed to eliminate such double-counting \cite{Manohar:2006nz}.  

We will address each of these issues without explicit calculation of the jet and soft loop integrals or use of an explicit infrared regulator. Instead  we just examine the regions of integration surviving in the sum over all relevant diagrams. We will work in pure dimensional regularization, and learn how to identify $1/\epsilon$ poles as infrared or ultraviolet in origin, clarifying the contribution made by scaleless integrals which are formally zero. We will thus conclude that the analysis is independent of the choice of any explicit IR regulator. In the process, we demonstrate the crucial role of zero-bin subtractions in obtaining physically-consistent, infrared-safe jet functions in angularity distributions for all $a<1$.  The ideas and methods illustrated through our discussion of angularity distributions are more generally applicable to other observables as well.

\section{Angularity Distributions in SCET}
\label{sec:SCET}

The factorization theorem for the angularity distributions $d\sigma/d\tau_a$ takes the form,
\begin{equation}
\label{factorizationtheorem}
\begin{split}
\frac{1}{\sigma_0}\frac{d\sigma}{d\tau_a} = H(Q;\mu)\! \int  &d\tau_a^n\, d\tau_a^\bn\, d\tau_a^s\, \delta(\tau_a-  \tau_a^n - \tau_a^\bn - \tau_a^s) \\
&\times J_a^n(\tau_a^n;\mu) J_a^\bn(\tau_a^\bn;\mu) S_a(\tau_a^s;\mu)\,,
\end{split}
\end{equation}
where $\sigma_0$ is the total $e^+e^-\to q\bar q$ Born cross-section, $H$ is a hard function given in the effective theory by the square of a matching coefficient dependent only on short-distance effects, $J_a^{n,\bn}$ are jet functions dependent on the partonic branching and evolution of each of the two back-to-back final state jets, and $S_a$ is a soft function dependent on the low energy radiation from the jets and the color exchange between them. All the functions depend on the factorization scale $\mu$, with this dependence cancelling in the full cross-section. The factorization theorem \eq{factorizationtheorem} for angularity distributions has been proved in full QCD \cite{Berger:2003iw} and in SCET \cite{Bauer:2008dt,Lee:2006nr}, for $a<1$, where this condition was derived from  the size of power corrections induced by replacing the thrust axis implicit in \eq{tauadef} with the collinear jet axis $\vect{n}$ \cite{Bauer:2008dt,Berger:2003iw}.  Our attempt to calculate perturbatively the jet and soft functions in \eq{factorizationtheorem} will provide a complementary way to deduce this condition and an intuitive explanation of the absence of infrared divergences in the jet and soft functions for $a<1$ and their appearance for $a\geq 1$.

Collinear and soft modes in SCET are distinguished by the scaling of the momenta of the particles they describe. The light-cone components $p = (n\cdot p,\bn \cdot p,p_\perp)$ of collinear modes, where $n,\bn$ are light-cone vectors in the $\pm z$ directions, scale as $Q(\lambda^2,1,\lambda)$ or $Q(1,\lambda^2,\lambda)$, and soft modes as $Q(\lambda^2,\lambda^2,\lambda^2)$. $Q$ is the hard energy scale in the process being considered (here, the center-of-mass energy in $e^+ e^-$ collisions), and $\lambda$ is a small ratio of energy scales, here $\lambda = \sqrt{\Lqcd/Q}$. Collinear momenta $p_c$ are split into a ``label'' piece $\tilde p_c$ containing the order $Q$ and $Q\lambda$ momenta, and a ``residual'' piece $k_c$ all of whose components are order $Q\lambda^2$. A redefinition of the collinear fields through multiplication by soft Wilson lines decouples soft and collinear modes in the SCET Lagrangian to leading order in $\lambda$ \cite{Bauer:2001yt}. 

The soft function $S_a$ in \eq{factorizationtheorem} is defined by
\begin{equation}
S_a(\tau_a^s;\mu) = \frac{1}{N_C}\!\Tr\bra{0}\overline Y_{\bar n}^\dag (0)Y_n^\dag(0)\delta(\tau_a^s - \hat \tau_a^s)Y_n(0)\overline Y_{\bar n} (0)\ket{0} ,
\label{softfunctiondef}
\end{equation}
and the jet functions $J_a^{n,\bn}$ by
\begin{subequations}
\label{jetfunctions}
\begin{align}
J_a^n(\tau_a^n,\mu)  \left(\frac{\nslash}{2}\right)_{\alpha\beta}  &=  \frac{1}{N_C}\Tr \int\frac{dl^+}{2\pi}\int \! d^4 x\,\ee^{il\cdot x}  \label{njetfunction} \\
&\quad \times\bra{0}\chi_{n,Q}(x)_\alpha\delta(\tau_a^n  - \hat \tau_a^n)\bar\chi_{n,Q}(0)_\beta\ket{0} \nonumber
 \\
J_a^\bn(\tau_a^\bn,\mu) \left(\frac{\bnslash}{2}\right)_{\alpha\beta}  &= \frac{1}{N_C}\Tr \int\frac{dk^-}{2\pi}\int\! d^4 x\,\ee^{ik\cdot x}  \label{nbarjetfunction}  \\
&\quad\times \bra{0}\bar\chi_{\bar n,-Q}(x)_\beta\delta(\tau_a^{\bar n}  - \hat \tau_a^{\bar n})\chi_{\bar n,-Q}(0)_\alpha\ket{0}. \nonumber
\end{align}
\end{subequations}
The traces are over colors, the light-cone momenta are defined $l^+ = n\cdot l$ and $k^- = \bn\cdot k$,  and the subscripts $Q$ on the jet fields in \eq{jetfunctions} specify that they create jets with total label momenta $Qn/2$ and $Q\bn/2$ \cite{Bauer:2001ct}.
The soft Wilson line $Y_n$ in the soft function is the path-ordered exponential of soft gluons, 
\begin{equation}
Y_n(z) = P\exp\left[ig\int_0^\infty ds\,n\cdot A_s(ns + z)\right]\,,
\end{equation} 
and similarly for $\overline Y_\bn$, with the bar denoting the anti-fundamental representation. The fields $\chi_{n,\bn}$ in the jet function are the product of collinear Wilson lines and quarks, $\chi_n = W_n^\dag\xi_n$,
where 
\begin{equation}
W_n(z) = P\exp\left[ig\int_{0}^\infty ds\,\bn\cdot A_n(\bn s + z)\right]\,,
\end{equation}
and similarly for $W_\bn$. The operator $\hat\tau_a$ acts on final states $\ket{X}$ according to
\begin{equation}
\label{tauhat}
\hat\tau_a\ket{X} = \frac{1}{Q}\sum_{i\in X} \abs{\vect{p}_i^\perp}\ee^{-\abs{\eta_i}(1-a)}\ket{X}\,,
\end{equation}
and is constructed from the energy-momentum tensor $T_{\mu\nu}$ \cite{Korchemsky:1997sy}, and the operators $\hat\tau_a^{n,\bn,s}$ in \eqs{softfunctiondef}{jetfunctions} are constructed by keeping only the $n,\bn$-collinear or soft terms in $T_{\mu\nu}$ \cite{Bauer:2008dt}. For further details of the SCET Lagrangian and the Feynman rules, we refer the reader to Refs.~\cite{Bauer:2000yr,Bauer:2001ct,Bauer:2001yt}. 

Next we proceed to examine the infrared behavior of $\mathcal{O}(\as)$ contributions to  the soft and jet functions of \eqs{softfunctiondef}{jetfunctions} calculated in perturbation theory.

\section{Divergences in the Soft Function}

\begin{figure}[b]
\centerline{\resizebox{8.5cm}{!}{\includegraphics{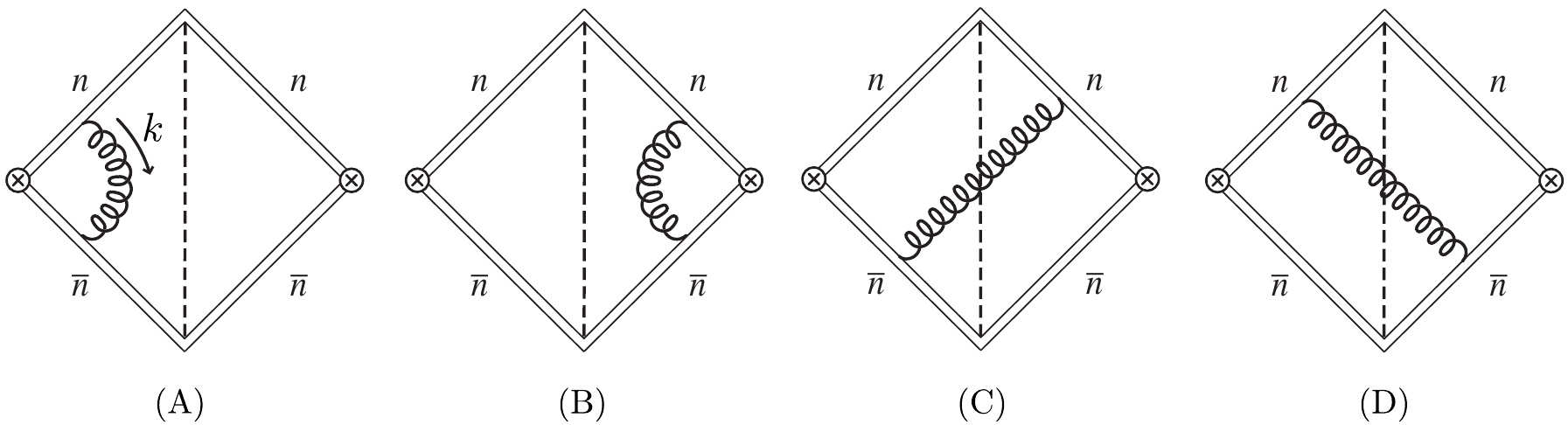}}}
\vspace{-.25cm}
{\caption[1]{The (A), (B) virtual and (C), (D) real gluon contributions to the soft function. The gluons all have momentum $k$.}
\label{softfunction} }
\end{figure}

 The soft function is calculated from the cut diagrams in Fig.~\ref{softfunction}, with an additional delta function $\delta(\tau_a^s-\tau_a(X_s))$ inserted along the cut, where $X_s$ is the final state created by the cut, and $\tau_a(X)$ is given by \eq{tauadef}. This modified cutting rule is required by the insertion of the $\delta(\tau_a^s-\hat\tau_a^s)$ operator in \eq{softfunctiondef} \cite{long}.

In diagrams (A) and (B) of Fig.~\ref{softfunction} with a virtual gluon, this delta function is just $\delta(\tau_a^s)$, whose coefficient is given by the virtual gluon loop integral. Using pure dimensional regularization in $d = 4-2\epsilon$ dimensions, this integral is scaleless and defined to be zero. This zero is actually a quantity proportional to $1/\eUV - 1/\eIR$, and ordinarily plays the role of cancelling $1/\eIR$ divergences in diagrams (C) and (D) in which the cut creates a real gluon, and converting them to $1/\eUV$ \cite{Manohar:1997qy,Manohar:2006nz}.

It is not at all obvious, however, how this cancellation can occur, since the virtual diagrams are independent of $a$ while the real gluon diagrams depend explicitly on $a$.  One often just prescribes the virtual diagrams to  take a form that converts the $1/\eIR$ poles in the real diagrams to $1/\eUV$, but this prescription is \emph{ad hoc} and, as we will see below, potentially misleading.

The soft function takes the general form 
\begin{equation}
S_a(\tau_a^s) = A\delta(\tau_a^s) + \sum_n B_n\left[\frac{\theta(\tau_a^s)\log^n\tau_a^s}{\tau_a^s}\right]_+\,.
\end{equation}
Since the virtual diagrams are proportional to $\delta(\tau_a^s)$, to study how they cancel the IR poles in the real diagrams, we only need to isolate the coefficient $A$ of $\delta(\tau_a^s)$.
By integrating all the diagrams over $\tau_a^s$ between 0 and 1, using the property of the plus functions $\int_0^1 dx [\theta(x)(\log^n x)/{x}]_+ = 0$,
we isolate $A$. We will denote as $I_S^V$ and $I_S^R$ respectively the virtual and real diagrams' contributions to $A$.

\begin{figure}[t]
\centerline{\resizebox{8.5cm}{!}{\includegraphics{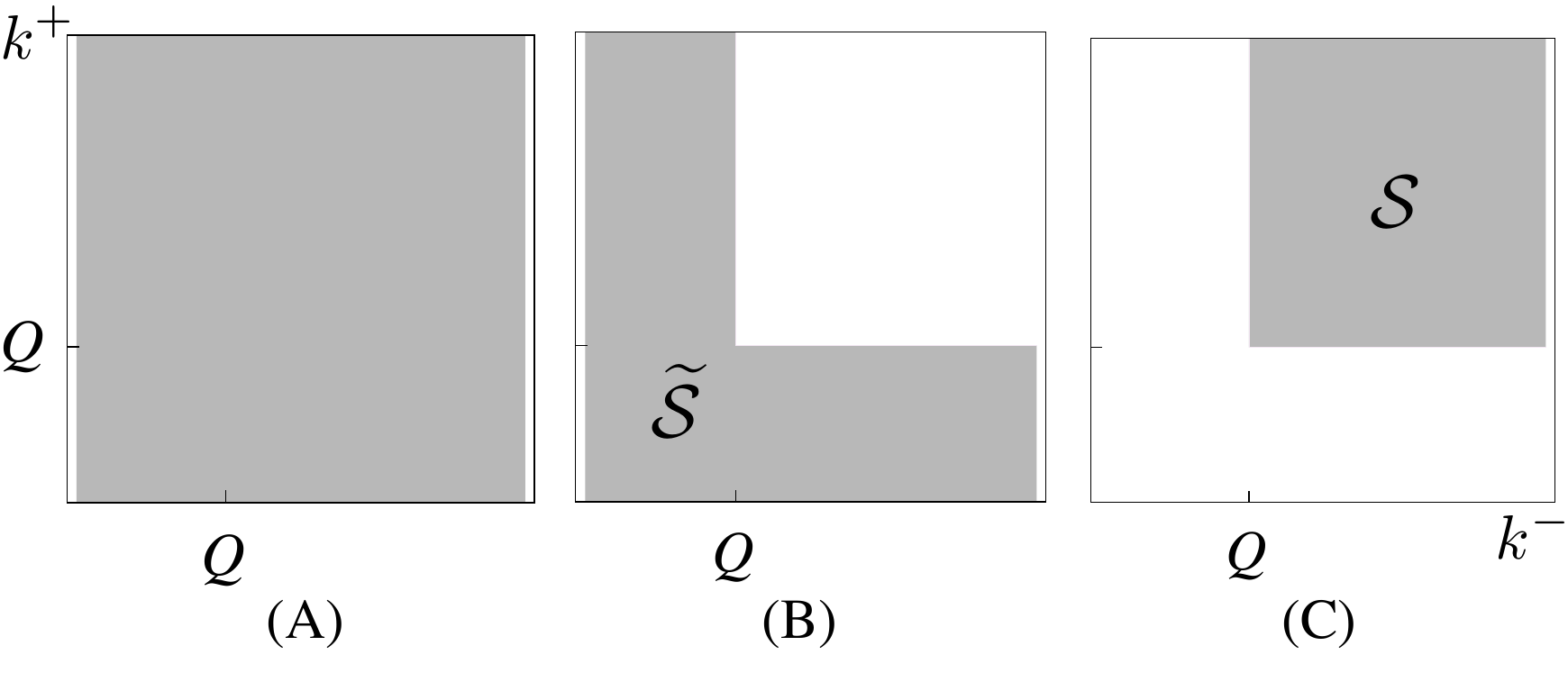}}}
\vspace{-.5cm}
{ \caption[1]{Regions of integration in the $k^-,k^+$ plane for the coefficient of $\delta(\tau_0^s)$ in the  $a=0$ soft function $S_0(\tau_0^s)$. (A) The region of integration for the virtual diagrams is the entire first quadrant. (B) For the real diagrams the region is $\widetilde{\mathcal{S}}$, which contains IR divergent regions. (C) These are converted in the sum of virtual and real graphs into the purely UV  region $\mathcal{S}$.
\label{softregionsa0}}}
\end{figure}

The virtual diagrams' contribution to $A$  is
\begin{align}
I_S^V = -\frac{\as \CF}{\pi} \frac{( 4\pi \mu^2 )^\epsilon}{\Gamma(1-\epsilon) } \int_0^\infty \! d k^+  \int_0^\infty d k^- (k^+ k^-)^{-1-\epsilon}\,,
\label{virtregion}
\end{align}
whose integration region is the entire first quadrant of the $k^\pm$ plane as shown in Fig.~\ref{softregionsa0}A.

In the real gluon diagrams, the cut creates a state $X_g$ with a single soft gluon, and the operator $\delta(\tau_a^n-\hat\tau_a^n)$ acting on $X_g$ introduces  the delta function  $\delta(\tau_a^s - \tau_a(X_g))$ into the integral over the gluon momentum $k$, where
\begin{align}
\tau_a(X_g)  =
\begin{cases}
& \frac{1}{Q} \abs{k^+}^{1-\frac{a}{2}} \abs{k^-}^{\frac{a}{2}} \quad {\rm for} \quad k^-\geq k^+ \\
&\frac{1}{Q} \abs{k^-}^{1-\frac{a}{2}} \abs{k^+}^{\frac{a}{2}} \quad {\rm for} \quad k^- < k^+\,.
\end{cases}
\label{taucalc}
\end{align}
  The real diagrams thus contribute
\begin{equation}
\label{realregion}
I_S^R = \frac{\as C_F}{\pi}\frac{(4\pi\mu^2)^\epsilon}{\Gamma(1-\epsilon)}\doubleint_{\widetilde{\mathcal{S}}} dk^+ dk^- (k^+ k^-)^{-1-\epsilon}
\end{equation}
to  $A$. 
$I_S^R$ depends explicitly on $a$ through the integration region determined by the delta function $\delta(\tau_a^s - \tau_a(X_g))$, restricting  gluon momenta $k$ to the  region $\widetilde{\mathcal{S}}$, given by $k^\pm > 0$ and 
\begin{align}
 & \qquad (k^-)^\frac{a}{2}(k^+)^{1-\frac{a}{2}} < Q \quad \text{for} \quad k^-\geq k^+ \nn\\
& \qquad (k^+)^\frac{a}{2}(k^-)^{1-\frac{a}{2}} < Q \quad \text{for} \quad  k^- < k^+
\,.\end{align}
This region is plotted in \fig{softregionsa0}B for $a=0$. 

The virtual and real integrals $I_S^{R,V}$ contain exactly the same integrand, but with opposite relative signs and integrated over different regions of the $k^\pm$ plane. Thus, in the sum of the virtual and real integrals $I_S^R + I_S^V$, the integrals over the overlapping part of the regions cancel, leaving an integral over the region $\mathcal{S}$, as  illustrated in Fig.~\ref{softregionsa0}C. 
\begin{figure}[t]
\vspace{-.125cm}
\centerline{\resizebox{8.5cm}{!}{\includegraphics{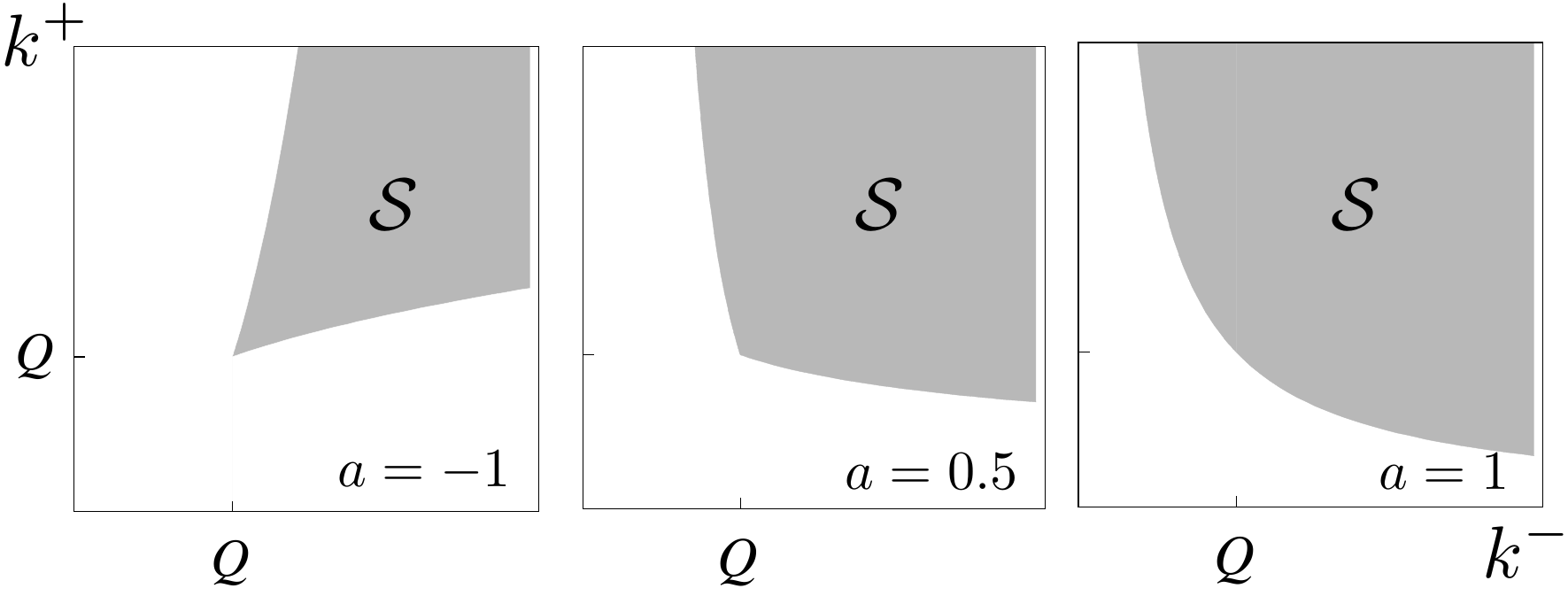}}}
\vspace{-.37cm}
{ \caption[1]{Region of integration in the $k^-,k^+$ plane for the coefficient of $\delta(\tau_a^s)$ in the soft function $S_a(\tau_a^s)$ for $a=-1$, $a=0.5$, and $a=1$. The region $\mathcal{S}$ is formed by summing real and virtual diagram regions as in Fig.~\ref{softregionsa0}. For $a<1$, $\mathcal{S}$ always remains above the line $k^+ k^- = Q^2$, which is the boundary of the $a=1$ region and divides the infrared and ultraviolet regions of the soft loop integration.
\label{softregions}}}
\end{figure}

In \fig{softregions} we plot the integration regions resulting from the sum of virtual and real diagrams for several other values of $a$.
For $a\leq 0$, the resulting region of integration $\mathcal{S}$ always satisfies $k^\pm\geq Q$, and is manifestly a purely UV region.  Between $0<a<1$, $k^\pm$ do approach zero on the boundaries, but the product $k^+ k^- \sim \vect{k}_\perp^2$ always is greater than $Q^2$, so  divergences in the soft loop integral are still purely UV. For $a=1$, the boundary of the region is the line of constant $k^+ k^- = Q^2$. For $a>1$, the boundary drops below this line, so, as $k^\pm \to \infty$, the product drops to $k^+ k^-\sim\vect{k}_\perp^2 \rightarrow 0$ in the region along the boundary. But such momenta are in fact collinear. The sum of soft diagrams still contains contributions from collinear modes. Continuing to explicitly evaluate $I_S^R + I_S^V$, we find this sum of integrals is convergent for $a<1$ when $\epsilon>0$, so the poles are $1/\eUV$, but not for $a\geq 1$, in which case uncancelled IR divergences remain. 

The shapes of the regions in Figs.~\ref{softregionsa0} and \ref{softregions} also tell us how using an explicit IR regulator would affect our analysis, and in fact teaches us that the choice of regulator must be made with care. For example, we might choose an effective cutoff $\lambda$ on $k^\pm$ in soft loop integrals as used in \cite{Chay:2004zn}, in which  the soft function for jet energy distributions was calculated to one loop and argued to be IR finite. In these cases this regulator successfully cuts off the divergences arising from the regions $k^\pm\to 0$.  However, using this regulator for $0<a<1$, we find that the soft function still contains $\log\lambda$ divergences and $1/\eUV$ divergences even though the above analysis shows that $I_S^R + I_S^V$ is actually IR finite. From Fig.~\ref{softregions} it is evident that a lower cutoff on $k^\pm$ also cuts off regions where $k^\pm\rightarrow\infty$, so it acts also partially as a UV cutoff. This underscores the challenge of defining consistent, explicit IR regulators in SCET 
\cite{Bauer:2003td,Chiu:2009yx}.

We draw two lessons from the analysis thus far. The first is that in pure dimensional regularization, the coefficient of $(1/\eUV-1/\eIR)$ in a virtual diagram cannot be determined from the virtual diagram alone, but only together with the real diagram whose IR divergence it is supposed to cancel (cf. \cite{Idilbi:2007ff}). The reason that the virtual subtraction can depend on $a$ even though by itself it is independent of $a$ is that the area of overlap between the integration regions of real and virtual diagrams depends on $a$. The second is that the presence or absence of IR divergences in the sum of all contributing loop integrals can (and should) be determined before completely evaluating the integral with a given IR regulator. Looking at the shape of the region of integration in momentum space  as above avoids confusion about the consistency of the regulator itself.

\section{Divergences in the Jet Function}

Now we analyze the jet functions \eq{jetfunctions} at $\mathcal{O}(\as)$ in perturbation theory. We will observe the same breakdown of infrared safety as $a\to 1$ due to the momentum regions beginning to include IR divergent regions. We will consider just the jet function $J_a^n(\tau_a^n)$; identical analysis applies to $J_a^{\bn}(\tau_a^\bn)$.  
\begin{figure}[b]
\centerline{\resizebox{8.5cm}{!}{\includegraphics{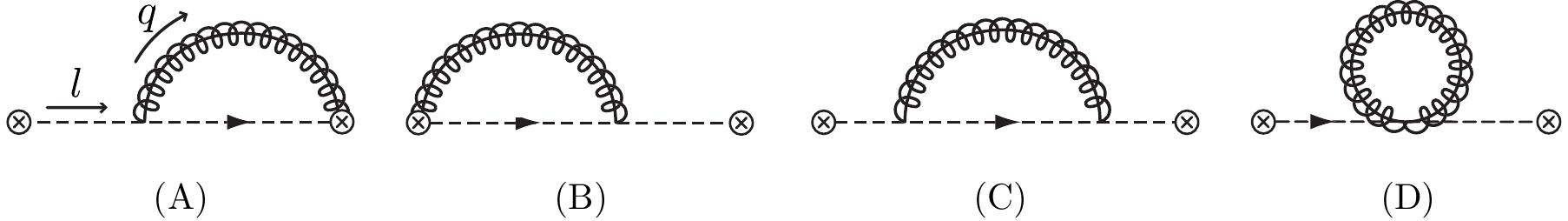}}}
\vspace{-.25cm}
{ \caption[1]{Diagrams contributing to the angularity jet
function $J_a^n(\taun)$. The total momentum through each graph is $Qn/2 + l$, and each gluon momentum is $q$. (A) Wilson line emission diagram and (B) its mirror; (C) sunset and (D) tadpole QCD-like diagrams. }
\label{jet-function} }
\end{figure}

The diagrams contributing to  $J_a^n(\tau_a^n)$ in \eq{njetfunction} are shown in Fig.~\ref{jet-function}. In graphs (A) and (B), the gluon is emitted from a collinear Wilson line $W_n$ or $W_n^\dag$ in the jet fields $\bar\chi_n,\chi_n$. The sum of graphs (C) and (D) is equivalent to graphs in full QCD by a field redefinition \cite{Becher:2006qw,Bauer:2008qu} and is manifestly IR finite for all $a<2$, and we will not consider them further here \cite{long}. 
The total momentum flowing through each diagram is $Qn/2 + l$, with the label component $Qn/2$ specified by the labels on the jet fields in the matrix elements in \eq{njetfunction}, and $l$ the residual momentum. The total momentum of the gluon in each loop is $q = \tilde q + q_r$, where $\tilde q$ is the  label momentum and $q_r$ the residual momentum. 
The diagrams must be cut in all possible places, and a delta function $\delta(\tau_a^n - \tau_a(X_n))$ inserted along each cut. To obtain $J_a^n(\taun)$ we then integrate over $l^+$ according to \eq{jetfunctions}.

An integral over the collinear gluon momentum $q$ is a sum over  $\tilde q$ and an integral over  $q_r$. The sum excludes the value  $\tilde q = 0$. The sum and integral can be combined into a single continuous integral over the total $q$ if a ``zero-bin subtraction'' is also taken \cite{Manohar:2006nz}, which avoids the double-counting of soft contributions between the soft and jet functions \cite{Berger:2003iw,Lee:2006nr,Chiu:2009yx,Idilbi:2007ff,Idilbi:2007yi}. Below we will refer to integrals or graphs before the zero-bin subtraction as ``na\"{\i}ve'', and after the subtraction ``collinear''.

Virtual graphs and zero-bin subtractions in pure dimensional regularization again contain scaleless integrals, which are zero, but as we observed in the calculation of the soft function, we must examine the regions of integration to observe how the cancellation of IR divergences among all the graphs occurs.

\begin{figure}[t]
\centerline{\resizebox{8.5cm}{!}{\includegraphics{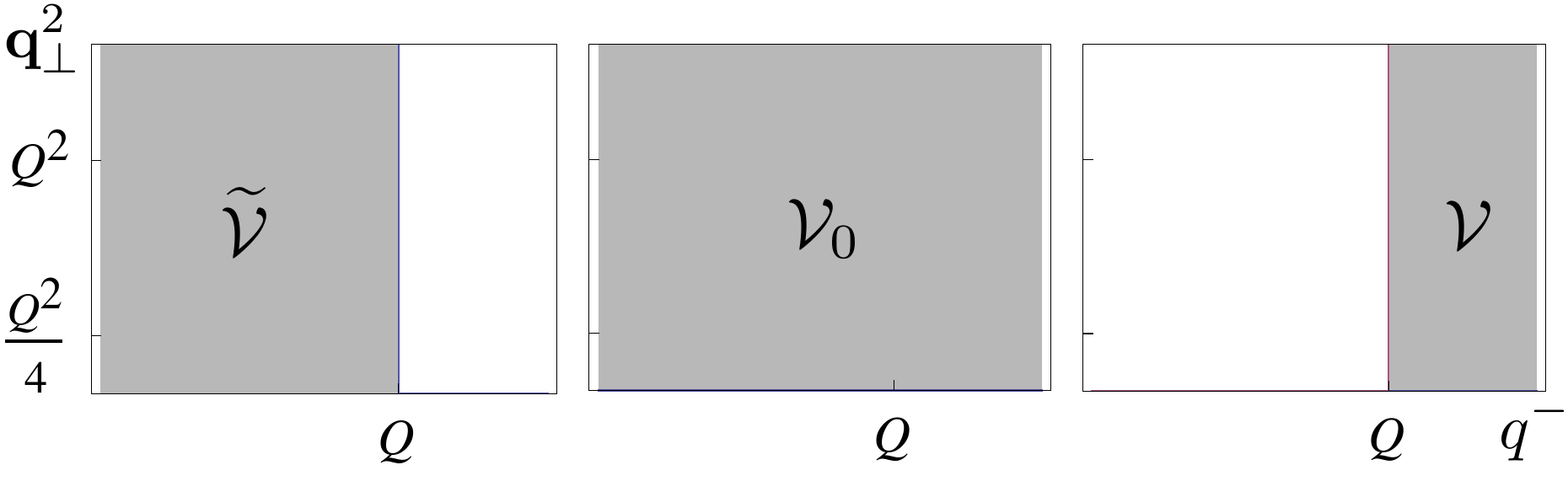}}}
\vspace{-.375cm}
{ \caption[1]{Regions of integration in the $q^-,\vect{q}_\perp^2$ plane for virtual gluon diagram contributions to the coefficient of $\delta(\tau_a^n)$ in the jet function $J_a^n(\tau_a^n)$. $\widetilde{\mathcal{V}}$ is the region for the na\"{\i}ve integral, $\widetilde{\mathcal{V}}_0$ for the zero-bin subtraction,  and $\mathcal{V}$ for the sum of these two contributions.
\label{virtualregions}}}
\end{figure}

Diagrams created by cuts through the single quark propagator  in graphs (A) and (B) in Fig.~\ref{jet-function} leave a virtual gluon  loop and  are proportional to $\delta(\tau_a^n)$, whose coefficient we extract. The na\"{\i}ve virtual graph contributes
\begin{equation}
\tilde I_{n}^V = -\frac{\as C_F}{\pi}\frac{(4\pi\mu^2)^\epsilon}{\Gamma(1-\epsilon)}\int_0^Q dq^- \int_0^\infty d\vect{q}_\perp^2 \frac{1}{(\vect{q}_\perp^2)^{1+\epsilon}}\left(\frac{1}{q^-} - \frac{1}{Q}\right) \,,
\end{equation}
which goes over region $\widetilde{\mathcal{V}}$ in Fig.~\ref{virtualregions}.
The zero-bin subtraction from the virtual graph is
\begin{equation}
I_{n0}^V = -\frac{\as C_F}{\pi}\frac{(4\pi\mu^2)^\epsilon}{\Gamma(1-\epsilon)}\int_0^\infty dq^- \int_0^\infty d\vect{q}_\perp^2 \frac{1}{(\vect{q}_\perp^2)^{1+\epsilon}}\frac{1}{q^-}\,,
\end{equation}
which goes over the whole first quadrant, region $\mathcal{V}_0$ in Fig.~\ref{virtualregions}.
So  the total virtual collinear contribution $I_n^V = \tilde I_n^V - I_{n0}^V$ is
\begin{equation}
I_n^V  = \frac{\as C_F}{\pi}\frac{(4\pi\mu^2)^\epsilon}{\Gamma(1-\epsilon)}\int_0^\infty d\vect{q}_\perp^2 \frac{1}{(\vect{q}_\perp^2)^{1+\epsilon}} \Biggl[\int_Q^\infty \frac{dq^-}{q^-} + \int_0^Q \frac{dq^-}{Q}\Biggr]\,,
\end{equation}
where the $1/q^-$ term is integrated over region $\mathcal{V}$  in Fig.~\ref{virtualregions} and the $1/Q$ term over $\widetilde{\mathcal{V}}$.

Now we add the contribution of the graphs (A) and (B) in Fig.~\ref{jet-function} cutting through the gluon loop, creating a final state $X_{qg}$ with a collinear quark and gluon, with
\begin{equation}
\tau_a(X_{qg}) = \frac{1}{Q} \left[(q^-)^{\frac{a}{2}}(q^+)^{1-\frac{a}{2}} +(Q \!-\!q^-)^{\frac{a}{2}}(l^+ \!-\! q^+)^{1-\frac{a}{2}}\right] \,
\end{equation}
and insert $\delta(\tau_a^n - \tau_a(X_{qg}))$ into the integral over the gluon momentum $q$. As in the case of the soft function, we need only to isolate the coefficient of $\delta(\tau_a^n)$ in the jet function $J_a^n(\tau_a^n)$ to study the cancellation of IR divergences with the virtual graphs. We do so by again integrating over $0<\tau_a<1$. 

\begin{figure}[t]
\centerline{\resizebox{8.5cm}{!}{\includegraphics{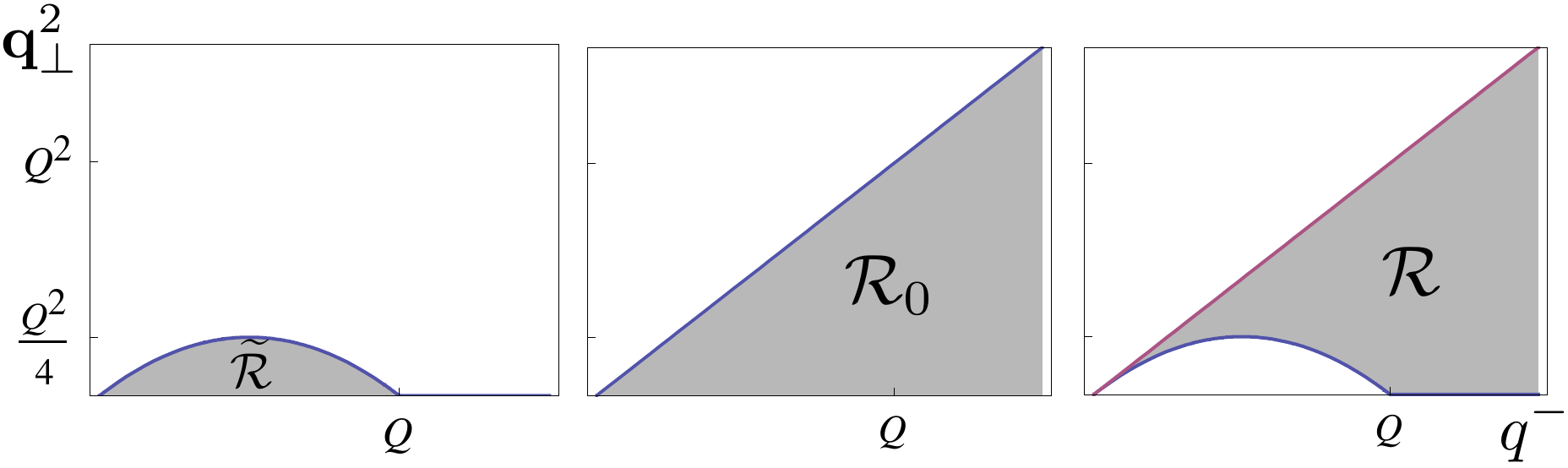}}}
\vspace{-.375cm}
{ \caption[1]{Regions of integration in the $q^-,\vect{q}_\perp^2$ plane for real gluon diagram contributions to the coefficient of $\delta(\tau_0^n)$ in the $a=0$ jet function $J_0^n(\tau_0^n)$. $\widetilde{\mathcal{R}}$ is the region for the na\"{\i}ve integral, $\widetilde{\mathcal{R}}_0$ for the zero-bin subtraction,  and $\mathcal{R}$ for the sum of these two contributions.
\label{realregions}}}
\end{figure}

The contribution to the coefficient of $\delta(\tau_a^n)$ from the na\"{\i}ve Wilson line graphs (A) and (B) in Fig.~\ref{jet-function} with a cut through the gluon loop is
\begin{equation}
\label{naivecollinear}
\tilde I_n^R = \frac{\as C_F}{\pi}\frac{(4\pi\mu^2)^\epsilon}{\Gamma(1-\epsilon)}\doubleint_{\widetilde{\mathcal{R}}}dq^- d\vect{q}_\perp^2 \frac{1}{(\vect{q}_\perp^2)^{1+\epsilon}}\left(\frac{1}{q^-} - \frac{1}{Q}\right)\,,
\end{equation}
where $\widetilde{\mathcal{R}}$ is the region in the first quadrant of the $q^-,\vect{q}_\perp^2$ plane under the curve 
\begin{equation}
\vect{q}_\perp^2 = \left\{Q \left[\frac{1}{(Q-q^-)^{1-a}} + \frac{1}{(q^-)^{1-a}}\right]^{-1}\right\}^{\frac{1}{1-a/2}}\,,
\end{equation}
shown for $a=0$ in Fig.~\ref{realregions}. The zero-bin subtraction is 
\begin{equation}
I_{n0}^R =   \frac{\as C_F}{\pi}\frac{(4\pi\mu^2)^\epsilon}{\Gamma(1-\epsilon)}\doubleint_{\mathcal{R}_0} \!\! dq^- d\vect{q}_\perp^2 \frac{1}{(\vect{q}_\perp^2)^{1+\epsilon}}\frac{1}{q^-}\,,
\end{equation}
where $\mathcal{R}_0$ is the region given by $q^->0$ and 
\begin{equation}
0<\vect{q}_\perp^2 < \bigl[Q (q^-)^{1-a}\bigr]^{\frac{1}{1-a/2}}\,,
\end{equation}
shown for $a=0$ in Fig.~\ref{realregions}.
Subtracting the two integrals, $I_n^R = \tilde I_n^R - I_{n0}^R$, yields the correct collinear integral,
\begin{equation}
\label{InR}
\begin{split}
I_n^R = -\frac{\as C_F}{\pi}\frac{(4\pi\mu^2)^\epsilon}{\Gamma(1-\epsilon)}\Biggl[\doubleint_{\mathcal{R}} & dq^- d\vect{q}_\perp^2 \frac{1}{(\vect{q}_\perp^2)^{1+\epsilon}}\frac{1}{q^-} \\
+ \doubleint_{\widetilde{\mathcal{R}}} & dq^- d\vect{q}_\perp^2 \frac{1}{(\vect{q}_\perp^2)^{1+\epsilon}}\frac{1}{Q}\Biggr] \,.
\end{split}
\end{equation}
where the first integral goes over the region $\mathcal{R}$ formed by removing $\widetilde{\mathcal{R}}$ from $\mathcal{R}_0$, illustrated in Fig.~\ref{realregions} for $a=0$.
The second integral, containing $1/Q$, still goes over $\widetilde{\mathcal{R}}$.

\begin{figure}[t]
\centerline{\resizebox{8.5cm}{!}{\includegraphics{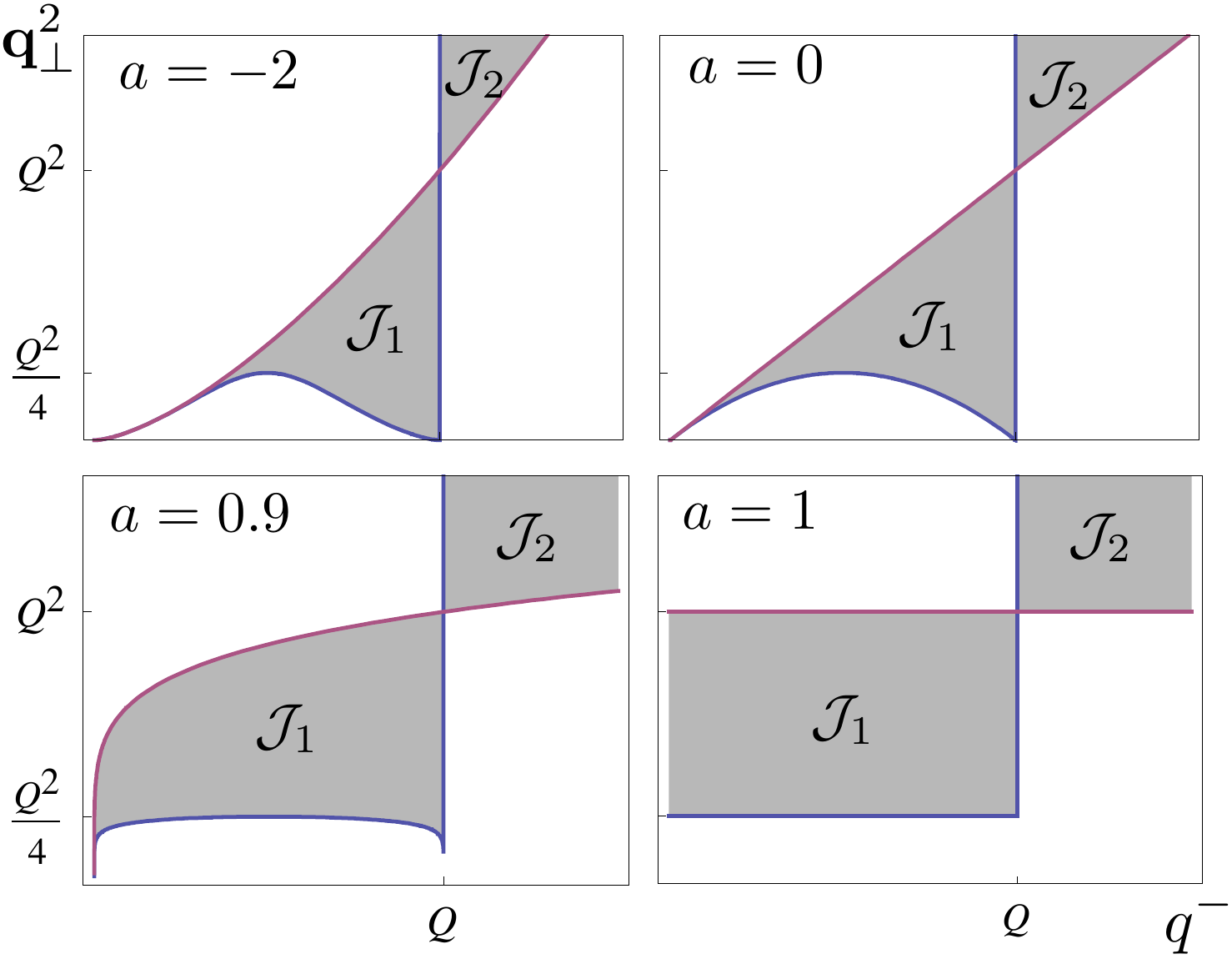}}}
\vspace{-.375cm}
{ \caption[1]{Regions of integration formed by combining real and virtual diagram regions in Figs.~\ref{virtualregions} and \ref{realregions}. The regions $\mathcal{J}_{1,2}$ result from subtracting the virtual diagram region $\mathcal{V}$ in Fig.~\ref{virtualregions} from the real diagram region $\mathcal{R}$ in Fig.~\ref{realregions}. For $a=1$, the region encounters an unregulated divergence at $q^-=0$. 
\label{jetregions}}}
\end{figure}

Thus, the sum $I_n = I_n^V + I_n^R$ of the collinear virtual and real graphs is
\begin{equation}
\label{totalIn}
\begin{split}
I_n = -\frac{\as C_F}{\pi}\frac{(4\pi\mu^2)^\epsilon}{\Gamma(1-\epsilon)}\Biggl[\doubleint_{\mathcal{J}_1-\mathcal{J}_{2}} \!\!  &dq^- d\vect{q}_\perp^2 \frac{1}{(\vect{q}_\perp^2)^{1+\epsilon}}\frac{1}{q^-} \\
- \doubleint_{\widetilde{\mathcal V} - \widetilde{\mathcal{R}}} \!\!  &dq^- d\vect{q}_\perp^2 \frac{1}{(\vect{q}_\perp^2)^{1+\epsilon}}\frac{1}{Q} \Biggr]\,,
\end{split}
\end{equation}
where the regions $\mathcal{J}_{1,2}$ are shown in Fig.~\ref{jetregions} for several values of $a$, and  the notation $\mathcal{J}_1 - \mathcal{J}_2$ means the integrand has an extra minus sign in $\mathcal{J}_2$. $\mathcal{J}_2$ contains only UV divergences. For all $a<1$, $\mathcal{J}_1$ avoids the boundary at $q^- = 0$, and the integral is convergent for $\epsilon>0$. The $1/\epsilon$ poles in this integral, as well as in the integral on the second line of \eq{totalIn}, are then purely UV. For $a=1$, however, the region $\mathcal{J}_1$ reaches the boundary at $q^- =0$, and the integral is no longer finite. The jet function is not infrared safe for $a\geq 1$, just as we found for the soft function.

Although the full distribution $d\sigma/d\tau_a$ is infrared safe for $a<2$, for $a\geq 1$, contributions of the soft and collinear modes of SCET with the momentum scalings specified in Sec.~\ref{sec:SCET} (so-called $\SCETa$ modes) do not entirely separate from each other. The soft integration regions illustrated in \fig{softregions} for $a\leq 1$ grow for $a>1$ to include the contribution of collinear modes, and the collinear integration regions in Fig.~\ref{jetregions} grow to include modes which are soft. Angularity distributions with $a\geq 1$ are dominated by jets so narrow that  collinear and soft modes have the same virtuality of order $\Lqcd$. We observe this in a full calculation of the jet and soft functions to $\mathcal{O}(\as)$ \cite{long}, which manifests the natural scales in the jet and soft functions where large logarithms are minimized, $\mu_J = Q\tau_a^{1/(2-a)}$ and $\mu_S = Q\tau_a$, which become equal at $a=1$. Thus, the separation of scales required by the formalism of \SCETa no longer holds. In this case, the modes may be distinguished by their rapidity, as was proposed in the formalism of \SCETb \cite{Manohar:2006nz}.

\section{Conclusions}

Although hard-scattering cross-sections can be written formally in  a factorized form based on na\"{\i}ve application of a formalism such as SCET, the properties of the chosen observable determine whether or not the effective theory is applicable and, so, whether the factorization theorem is actually valid. Such a theorem must pass a number of tests. The method we have presented is a straightforward and intuitive test of the infrared-safety of jet and soft functions in a proposed factorization theorem. We illustrated the method at $\mathcal{O}(\as)$ for angularities, whose tunable parameter $a$ allowed us to study the continuous progression from infrared-safety of jet and soft functions for $a<1$ to its breakdown for $a\geq 1$, but the method is more generally applicable to other observables as well. The test will reveal those observables for which the na\"{\i}ve \SCETa factorization fails. Through our analysis, we have illustrated the crucial role of zero-bin subtractions in effective field theory, and the manner in which scaleless integrals in pure dimensional regularization convert IR into UV divergences in infrared-safe quantities, without choosing any \emph{ad hoc} prescriptions, allowing one to classify IR and UV divergences independently of an explicit IR regulator.

\section*{Acknowledgements}
We are grateful to C. Bauer for many valuable discussions, constant encouragement, and extensive feedback on the draft. We also thank I. Stewart for useful discussions, and B. Lange and Z. Ligeti for careful review of the draft. CL and GO are grateful to the Institute for Nuclear Theory at the University of Washington for its hospitality during a portion of this work. AH is supported in part by an LHC Theory Initiative Graduate Fellowship. This work was supported in part by the U.S. Department of Energy under Contract DE-AC02-05CH11231, and in part by the National Science Foundation under grant  PHY-0457315.

\end{document}